\documentclass[prl,letterpaper,twocolumn,showpacs,superscriptaddress,amsmath,amsfonts,amssymb,preprintnumbers]{revtex4}

\usepackage[dvips]{graphicx}
\usepackage[dvips]{graphics}
\usepackage{times}

\begin{document}

\title{2D orbital-like magnetic order in ${\rm La_{2-x}Sr_xCuO_4}$} 

\author{V. Bal\'edent} 
\affiliation{Laboratoire L\'{e}on Brillouin, CEA-CNRS, CEA Saclay, 91191 Gif-sur-Yvette Cedex, France}

 \author{B. Fauqu\'e} 
\affiliation{Laboratoire L\'{e}on Brillouin, CEA-CNRS, CEA Saclay, 91191 Gif-sur-Yvette Cedex, France}

\author{Y.~Sidis}
\affiliation{Laboratoire L\'{e}on Brillouin, CEA-CNRS, CEA Saclay, 91191 Gif-sur-Yvette Cedex, France}

\author{N.B. Christensen}
\affiliation{ETH Zurich and Paul Scherrer Institute, CH--5232 Villigen PSI, Switzerland}

\author{S.~Pailh\`{e}s}, 
\affiliation{Laboratoire L\'{e}on Brillouin, CEA-CNRS, CEA Saclay, 91191 Gif-sur-Yvette Cedex, France}
\affiliation{ETH Zurich and Paul Scherrer Institute, CH--5232 Villigen PSI, Switzerland}

\author{K. Conder}
\affiliation{ETH Zurich and Paul Scherrer Institute, CH--5232 Villigen PSI, Switzerland}

 \author{E. Pomjakushina} 
\affiliation{ETH Zurich and Paul Scherrer Institute, CH--5232 Villigen PSI, Switzerland}

\author{J. Mesot}
\affiliation{ETH Zurich and Paul Scherrer Institute, CH--5232 Villigen PSI, Switzerland}

\author{P.~Bourges$^\ast$}
\affiliation{Laboratoire L\'{e}on Brillouin, CEA-CNRS, CEA Saclay, 91191 Gif-sur-Yvette Cedex, France}

\pacs{}
\begin{abstract} 
In high temperature copper oxides superconductors, a novel magnetic order associated with the pseudogap phase has been identified in two different cuprate families over a wide region of temperature and doping. We here report the observation below 120 K of a similar magnetic ordering in the archetypal cuprate ${\rm La_{2-x}Sr_xCuO_4}$ (LSCO) system for x=0.085.  In contrast to the previous reports, the magnetic ordering in LSCO is {\it\bf only} short range with an in-plane correlation length of $\sim$ 10 \AA\ and is bidimensional (2D). Such a less pronounced order suggests an interaction with other electronic  instabilities. In particular, LSCO also exhibits a strong tendency towards stripes ordering at the expense of the superconducting state.
\end{abstract}

\maketitle

The origin of the pseudogap phase remains one of the most animated debate in the high temperature copper oxides superconductors. Occuring in the normal state and over a wide region of doping, the pseudogap is visible in both magnetic and charge properties of all cuprates superconductors\cite{timusk,revue}. Using polarized neutron diffraction, a novel long range magnetic order has been recently established in two different cuprate families ${\rm YBa_2Cu_3O_{6+x}}$ (YBCO) \cite{fauque,mook} and ${\rm HgBa_2CuO_{4+\delta}}$  (Hg1201)\cite{li}. This observation is of primary importance, since the temperature of this magnetic transition matches the one of the pseudogap regime, providing a strong support in favor of a true phase transition \cite{revue,varmaQCP}. While the novel magnetic order also breaks time reversal symmetry, it should be described as a  {\bf Q}=0 antiferromagnetic order (AFO), {\it i.e.} a magnetic order preserving the translation symmetry of the lattice (TSL), but with a staggered magnetic pattern present within each unit cell. This magnetic order has been predicted in the circulating current theory of the pseudogap state\cite{varma}. That suggests that this {\bf Q}=0 AFO corresponds to an orbital-like order. In this theory, the magnetic moments are indeed associated with two opposite orbital moments per unit cell generated by closed current loops\cite{varma}. Recent variational Monte-Carlo calculations in extended Hubbard model show that orbital moments can indeed develop once the apical oxygen orbitals are taken into account\cite{weber}. However, the orbital nature of the observed magnetic moments is still an open experimental issue. Alternativelty, spin-based models could actually describe the observed magnetic peaks. For instance, a model of opposite spins on the oxygen sites has been proposed\cite{fauque}. 

In the archetypal HTS ${\rm La_{2-x}Sr_xCuO_4}$ (LSCO), the low energy spin excitation spectrum is dominated by incommensurate (IC) spin fluctuations around the planar antiferromagnetic (AF) wavevector at $\rm Q_{IC}= Q_{AF} \pm (\delta,0) \equiv Q_{AF}\pm (0,\delta)$ \cite{yamada} with $\rm Q_{AF}=(1/2,1/2)$. In isotructural compounds ${\rm La_{2-x}Ba_xCuO_4}$ \cite{fujita} and ${\rm (La,Nd)_{2-x}(Sr,Ba)_xCuO_4}$ \cite{tranquada-95}, where bulk superconductivity is strongly reduced, spin (SDW) and charge (CDW) density wave orders develop respectively at $\rm Q_{IC}$ and $\rm 2Q_{IC}$ \cite{tranquada}. Furthermore, IC spin excitations are also observed in strongly underdoped ${\rm YBa_2Cu_3O_{6.45}}$  \cite{hinkov}. These fluctuations indicate that the rotation invariance of the system is spontanously broken below 150 K. All of these properties can be understood within the charge stripe model \cite{stripes}, for instance. Stripes can be viewed as a filamentary charge organisation, separating AF domains in anti-phase. When charge stripes appear, but still fluctuate, the rotation symmetry is first broken. When fluctuating stripes are pinned down on the lattice or defects, they become static and the resulting charge and spin order ultimately breaks the translation invariance. By analogy with liquid crystals \cite{stripes}, the two phases are often called nematic and smectic stripe phases.

While the {\bf Q}=0 AFO (or magnetic orbital-like order) has been observed in a broad doping and temperature range in YBCO and Hg1201 systems, the static stripe-like order essentially develops in LSCO either at low doping near the Mott-insulating state or near the locking-in composition of x=1/8 \cite{fujita,julien,mitrovic,chang} reaching its maximum temperature of 20 K\cite{julien}. We here report a study of  the {\bf Q}=0 AFO in one LSCO sample with 8.5 \% of Sr. At the difference of YBCO and Hg1201, the observed magnetic order is short-range with an in-plane correlation lengh which does not exceed 10 \AA.  The magnetic order is basically two-bidimensional (2D) with an intensity spreaded along c$^*$ perpendicular to the CuO$_2$ layers. Meanwhile, we measure the low energy incommensurate magnetic fluctuations, usually associated with dynamical stripes. They exhibit a marked change at 120 K, pointing out an unexpected interaction between both magnetic correlations that could be entangled. 

The sample consists of three single crystals (total mass 7g) obtained by the traveling solvent floating zone method similar to the ones reported by \cite{growth}, co-aligned within less than 1 deg. Most of the data have been obtained in a scattering plane where all Bragg peaks like $\bf{Q}$=(H,0,L) (in tetragonal notations for which a=b=3.82 \AA\ and c=13.15 \AA) were accessible. The superconducting transition $T_c$=22 K has been measured by both magnetic susceptibility on a small piece as well as using neutron depolarization on the whole sample, corresponding to a doping level of 8.5\% \cite{oda}. All the polarized neutron diffraction measurements were collected on the same 4F1 triple-axis spectrometer at the $Laboratoire\ L\acute{e}on\ Brillouin,\ Saclay,\ France$ as in previous measurements \cite{fauque,mook,li} with an incident neutron wavevector of $k_i=2.57$ \AA$^{-1}$. Although the actual symmetry of LSCO is orthorhombic, we are using here tetragonal notations for an easier comparison with other cuprates.

\begin{figure}[t]
\includegraphics[width=6.5 cm,angle=-90]{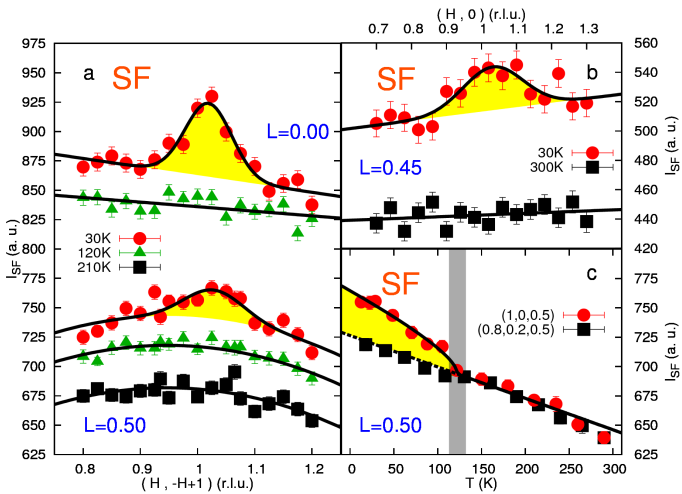}
\caption{Raw data in the spin-flip (SF) channel for a neutron polarisation {\bf H//Q}.
A) diagonal scans along the (1,-1) direction around Q=(1,0,L) at different temperatures for L=0 and L=0.5. B) longitudinal scans along the (1,0) direction around Q=(1,0,0.45).
C) Temperature dependence of neutron intensity (SF) at Q=(1,0,0.5) (red circles) and at a background position Q=(0.8,0.2,0.5) (black dots). For all figures, yellow areas represent the magnetic signal. The typical counting time is about 1 hour per point in order to get sufficient statistics.} 
\label{fig2}
\end{figure}


Respecting TSL, the magnetic moments of the {\bf Q}=0 antiferromagnetic  order scatter neutrons at the same positions as the Bragg peaks of the crystallographic structure. As a result, measurements have to be performed on the weakest nuclear Bragg peak having the proper symmetry for the magnetic phase in order to evidence small moments. As in other cuprates \cite{fauque,mook,li}, the Bragg point $\bf{Q}$=(1,0,1) offers the best compromise.  Attempts to observe a  long range order magnetic component on $\bf{Q}$=(1,0,1) was not succesfull in LSCO. Therefore, the {\bf Q}=0 antiferromagnetic long range order present in other cuprates is either absent in LSCO or too weak to be experimentally detected. One can give an upper estimate of less than 0.02 $\mu_B$ for a 3D ordered {\bf Q}=0 antiferromagnetic  moment in LSCO, as compared to the measured value of $\sim$ 0.1 $\mu_B$ in YBCO$_{6.6}$ for a doping $\sim$ 10 \% \cite{fauque}. 

\begin{figure}[t]
\includegraphics[width=6.5 cm,angle=-90]{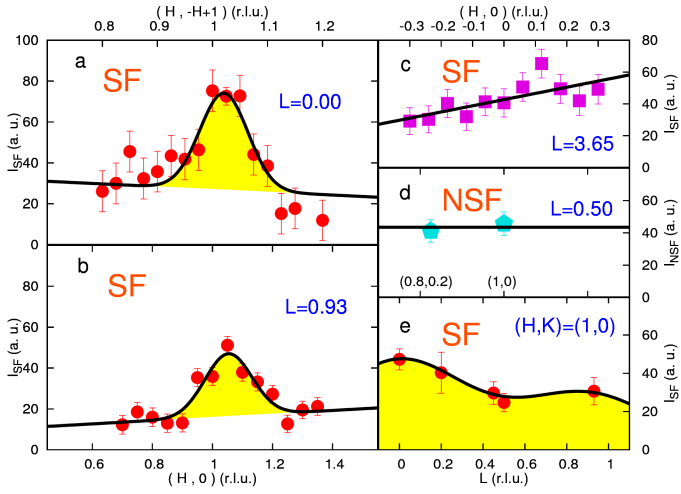}
\caption{Difference of  measurements performed at 30 K and 120 K for a neutron polarisation {\bf H//Q}.
A-C) Scans in the spin flip channel: A) Around {\bf Q}=(1,0,0) along a diagonal direction (1,-1), B) around {\bf Q}=(1,0,0.93) along a*.  
C) Around {\bf Q}=(0,0,3.65) along a*. D) Scan in the non spin flip channel around {\bf Q}=(1,0,0.5) along the diagonal direction (1,-1). E) L-dependence of the magnetic intensity (background substracted) determined from the various $Q$-scans measured across the {\bf Q}=(1,0,L) in the SF channel. The line across the points is a guide to the eye.
} 
\label{fig3}
\end{figure}

As pseudogap properties in LSCO are less accurate than in other cuprates\cite{timusk}, a more disordered state, characterized by finite correlation length, can be actually expected. In case of short range magnetic order, magnetic intensity would be redistributed in  momentum space, making its detection on top of a nuclear Bragg peak almost impossible even with polarized neutron diffraction. In order to look for a broader magnetic signal in Q-space, we then measure off the Bragg position but still localized around the same planar wavevectors, respecting the TSL, say $Q_{2D}=(1,0)$. Searching for a long or a short range order leads to different experimental issues.  In the former case, experimental difficulties are related to the neutron polarization stability at the Bragg peak. In the latter case, this is no more a problem as the signal is observed beside the Bragg position where the polarization stability is not a relevant issue.  Fig.\ref{fig2} shows scans in the spin-flip channel for {\bf H//Q} along either the diagonal (1,-1) direction (Fig.\ref{fig2}a) or along a* (Fig.\ref{fig2}b) across the wavevector {\bf Q}=(1,0,L) for L=0, 0.45 and 0.5. In all these scans, a peak is observed at 30 K and vanishes at high temperature. The scans along the diagonal direction show a signal centered at Q=(1,0,L) with a full width at half maximum (FWHM) of $\Delta_q= 0.11 \pm 0.02$ r.l.u independantly of L (the resolution FWHM is typically 0.07  r.l.u.). This leads to a correlation length of $\xi_{(1,1,0)} \equiv 2/\Delta_q \approx 11 \pm 2$\AA\ after resolution deconvolution. As shown by Fig.\ref{fig2}b, the scan along a* is broader, yielding $\xi_{(1,0,0)}= 8 \pm 3$ \AA. It is worth noticing that the signal/background ratio is only about 5\%. The Fig.\ref{fig2}c displays the temperature of the maximum of the signal at Q=(1,0) for L=0.5, as well as the temperature dependence of the background measured off the peak at Q=(0.8,0.2). The background exhibits a slope consistent with a Debye-Waller factor, on top of which the magnetic signal shows up on cooling down. The difference between both curves indicates a transition temperature  $T_{mag}$ around 120K, in agreement with high temperature scans shown on Fig.\ref{fig2}. Passing through $\rm T_c$, the magnetic intensity does not show any noticeable change (Fig.\ref{fig2}c). Finally, additional Q-scans at several L values have been also performed. Fig.\ref{fig3}a and b show difference of scans between 30 K and 120 K for L=0 and L=0.93. A magnetic peak is observed at any L indicating a quasi L-independent structure factor, as shown by Fig.\ref{fig3}e where we report the difference between the neutron intensity and the background from the scans measured at different L. The magnetic intensity is present at all measured L values. This implies that the magnetic correlations are basically two dimensional (2D).
 The observed magnetic signal is therefore a 2D short range order, occuring around the same symmetry points as in case of the long range  3D order in YBCO \cite{fauque,mook} and Hg1201 \cite{li}, corresponding to orbital-like magnetic order. The magnetic order is found to be static at the energy scale given by the spectrometer energy resolution which is about 1 meV. However, it might be fluctuating at lower energy.  

Fig.\ref{fig3}c shows the difference between 30 K and 120 K in the spin-flip channel and {\bf H//Q} for a scan along the a* axis around {\bf Q}=(0,0,3.65) which has been chosen such as the modulus of  {\bf Q} is similar to the wavevector {\bf Q}=(1,0,0.5). Interestingly, the  magnetic signal centered at (H,K)=(1,0) is absent for H=K=0. This result implies  a specific magnetic structure factor which needs more than one magnetic moment per unit cell and a sum of these moments equal to zero within each unit cell, as it is expected for the orbital magnetic order\cite{varma}. 

\begin{figure}[t]
\includegraphics[width=3.3 cm,angle=-90]{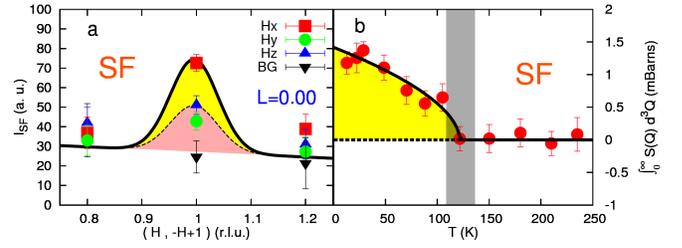}
\caption{A) Polarization analysis of the difference of scans measured at 30 K and 120 K in the spin-flip channel around {\bf Q}=(1,0,0) along a diagonal direction (1,-1). The neutron polarization is applied successively along three different directions ${\bf H_{\alpha}}$. The label $\alpha$ correspond to the Cartesian axis \{$ x,y,z$\}, so that the  $x$ axis is parallel to {\bf Q}, while the $z$ axis stands for the direction perpendicular to the scattering plane.  B) Temperature dependence of Q-integrated magnetic intensity, $S_{mag}$. 
}
\label{fig4}
\end{figure}

In a polarized neutron scattering measurements, the scattered intensity associated with a magnetic component simultaneaouly perpendicular to the momentum {\bf Q} and to the polarization direction shows up exclusively in the SF channel. As a result, the magnetic signal has to be purely spin flip for {\bf H//Q}. Accordingly, the temperature difference in the non-spin-flip channel for {\bf H//Q}  (Fig.\ref{fig3}d) shows no intensity peaked at H=1 in contrast to the corresponding SF data (Fig.\ref{fig2}a). Turning the neutron polarization direction, one can further single out the magnetic scattering associated with each magnetic component\cite{fauque,mook}.  The intensity measured for each polarisation is a sum of a magnetic intensity $I_{H \alpha}$ ($\alpha=\{ x,y,z\}$) and a background signal, which does not depend on the neutron polarisation. For only a magnetic scattering,  the neutron intensity should obey the following selection rule: $I_{Hx}=I_{Hy}+I_{Hz}$. 
The Fig.\ref{fig4}a shows the polarisation analysis of the scan along the diagonal for L=0. The expected relation for a magnetic scattering is observed  demonstrating the magnetic nature of this intensity centered at (H,K)=(1,0). Likewise, the neutron intensity for both polarizations perpendicular to {\bf Q}, each sensitive to either $M_z$ or $M_y$, has the same amplitude. Similar observations were made in the other cuprates \cite{fauque,mook,li}. This is typically what we would expect in case of a disordered magnetic state where the moments have no preferential direction. In case of an ordered magnetic state \cite{fauque,mook}, this may also indicate that the moments are pointing along a direction forming 45$^{o}$ angle between c* axis and the (a*,b*) plane. That could be understood within the orbital moments picture once apical oxygens are included\cite{weber} giving rise to a circulating current flowing on the faces of the CuO$_6$ octahedra.

In order to provide a more quantitative description of the observed magnetic signal,  we report on Fig. \ref{fig4}b the temperature dependence of the normalized Q-integrated magnetic structure factor, $S_{mag}\equiv\int d^3Q I_{mag}(Q)/\int d^3Q$. 
Firstly, the magnetic intensity at Q=(1,0,0.5) is obtained from the temperature dependence of the magnetic signal after subtraction of the background (both shown on Fig.\ref{fig2}c). The absolute value of  $S_{mag}$ is  then calculated from these data after integration in Q-space of the magnetic signal and after normalization of the magnetic intensity to the nuclear Bragg intensity. As represented in Fig. \ref{fig4}b, the deduced Q-integrated magnetic intensity, $S_{mag}$, reaches a value of 1.2 mbarns at the lowest temperature. A  magnetic local moment, $M_{loc}$, can be obtained from $S_{mag}$. The absolute value of the magnetic local moment is found around 0.1 $\mu_B$ at low temperature. Remarkably, similar amplitude for both the neutron structure factor and the magnetic moment has been reported for the {\bf Q}=0 AFO in both YBCO \cite{fauque,mook} and Hg1201 \cite{li}.


Here we have demonstrated that the orbital-like order is present in the archetypal LSCO system and is likely to be a generic property of superconducting cuprates.  In YBCO and Hg1201, the 3D long range order appears at a temperature $\rm T_{mag}$ matching the pseudogap temperature $\rm  T^{\star}$.  In our LSCO sample, the {\bf Q}=0 AF state settles in at $\rm T_{mag}$ $\sim$120 K.  As the doping is quite low (8.5~\%), a much larger $\rm  T^{\star}$ is typically expected from the generic phase diagram of high-Tc cuprates. However, it is worth pointing out that the features associated with the pseudogap temperature are less defined in LSCO than in the other cuprates \cite{timusk},  making it difficult to define $\rm  T^{\star}$  in an unequivocal way. Likewise, several anomalies have been reported close to $\rm T_{mag}$ in the specific heat  \cite{oda,ido}, the uniform spin susceptibility \cite{oda} and the Nernst effect \cite{ong} for LSCO samples in the same doping range: their interpretations should be reconsidered in light of our data. 

\begin{figure}[t]
\includegraphics[width=6.5 cm,angle=-90]{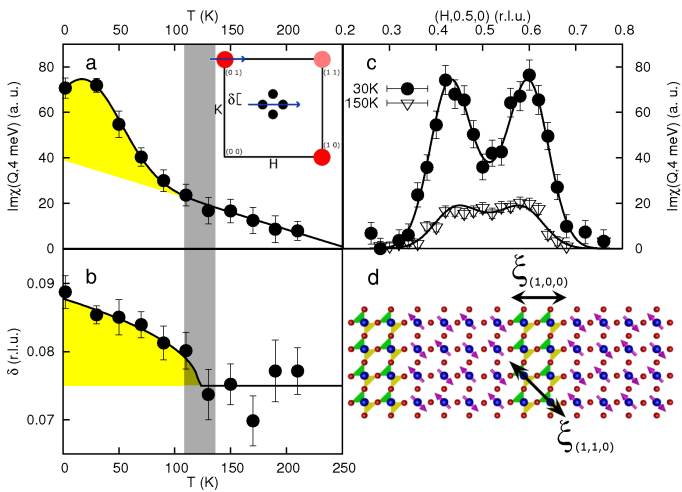}
\caption{ Unpolarized inelastic neutron scattering measurements of the IC magnetic fluctuations around 
$\rm Q_{IC}= Q_{AF} \pm (\delta,0)$: a) Temperature dependence  of the spin susceptibility at an energy of 4 meV: $
Im \chi (Q_{IC},\hbar\omega=4 {\rm meV})$. The inset represents the location of the different magnetic response in 
the a-b plane: {\bf Q}=0 AFO and IC spin fluctuations are shown in red and black, respectively.  
b) Temperature dependence of the IC parameter $\delta$. In  a) and b), the vertical dashed line indicates $\sim$ $\rm T_{mag}$. c) Typical H-scans across IC spin excitations at $\hbar\omega$=4 meV. The figure shows the imaginary part of the dynamical magnetic susceptibility $Im \chi (Q,\hbar\omega)$ at 30 K (full circles) and at 150 K (open triangles).  d) Schematic picture of the ${\rm CuO_2}$ plane  for a hole doping of 1/12 based on the bond centered stripes model discussed in ref. \cite{tranquada}.}
\label{fig5}
\end{figure}

In addition to the reduced value of  $\rm T_{mag}$, the {\bf Q}=0 AFO remains 2D and short range in LSCO. This frustration of the {\bf Q}=0 AF correlation could result from a competition with another electronic instability, namely the tendency toward stripes phase known, among cuprates, to occur specifically in the LSCO system. While the observed ordering temperature $\rm T_{mag} \sim$  120 K is larger than the smectic (i.e static) stripes-like ordering temperature ever reported in LSCO, {\it a priori}, there has to be a direct competition between the {\bf Q}=0 AFO  and the nematic (i.e dynamic) stripe phase as these phases are breaking different symmetries. To look for such a connection, we have studied the temperature dependence of the IC spin fluctuations  \cite{lipscombe,hirata}, usually associated with dynamical stripes at low energy. Our study of IC magnetism in our sample indicates strong dynamic IC fluctuations but no static IC magnetic peaks have been evidenced down to 1.5K. Typical scans across the incommensurate peaks at $\hbar\omega$=4 meV are shown in Fig. \ref{fig5}.c, indicating IC spin excitation at the wavevectors $\rm Q_{IC}$ with $\delta=0.085 \pm 0.005$ at T=30K (see Fig. \ref{fig5}.b). 
We discover that the incommensurability parameter $\delta$ exhibits a clear enhancement at $\rm T_{mag}$ (Fig. \ref{fig5}.b) which is accompagnied by an increase of the peak magnitude  (Fig. \ref{fig5}.a). Both behaviours have been actually already reported in stripe ordered Ba doped cuprate\cite{fujita} around 65 K. Here, we are able to relate these variations with the onset of the {\bf Q}=0 AFO at $\rm T_{mag}$, highlighting the relation between both electronic instabilities.

Recently, it has been shown  that orbital currents could also develop in doped two-leg spin ladders \cite{gabay}. Inspired by this work, a simple picture can emerge based on the bond centered stripes model\cite{tranquada}. Hole-poor regions exhibiting fluctuating copper spins (in purple in Fig. \ref{fig5}.d) are separated by hole-rich regions, depicted as hole doped 2-leg ladders. The {\bf Q}=0 AFO might develop within the charge stripes: the circulating current phase $\Theta_{II}$\cite{varma} is represented for a sake of example. The observed correlation lengths  of the {\bf Q}=0 AFO are shown along the (100) and (110) directions. In order to preserve the lattice translation invariance, magnetically order charge stripes should remain magnetically  decoupled from each other.
Excluded from the  hole poor AF domains where the spin fluctuations are growing at low temperature, the {\bf Q}=0 AFO 
correlation lengths are found limited by the size of the charge stripes (see Fig. \ref{fig5}.d). Still within this picture, the absence of 3D {\bf Q}=0 AFO in LSCO could be explained as stripes alternate directions as one goes from one ${\rm CuO_2}$ plane to the next.

Finally, at the same time as dynamical stripes are reported through incommensurate magnetic fluctuations, LSCO exhibits a short range {\bf Q}=0 AFO (or orbital-like magnetic order) which could be confined within the charge stripes. Having the same symmetry as the long range magnetic order reported in two others cuprates\cite{fauque,mook,li}, the nature of this magnetism, involving both copper and nearest neighbour oxgens, is likely a keystone towards understanding the physics of HTS.


{\bf Acknowledgements} We wish to thank T. Giamarchi, M. Greven, V. Hinkov, M.-H. Julien,  S.K. Kivelson, Yuan Li, D. Poilblanc, J.M. Tranquada, C.M. Varma and K. Yamada for interesting discussions and pertinent comments.
 This work was partly supported by NCCR MaNEP Project.
\normalsize{$^\ast$To whom correspondence should be addressed; E-mail: philippe.bourges@cea.fr}


\newpage


\begin{thebibliography}{10}


\bibitem{timusk}
T. Timusk,  and B. Statt,
 {\it Rep. Prog. Phys.} {\bf 62},  61 (1999).

\bibitem{revue} 
M.R. Norman and C. P\'epin,  
 {\it Rep. Prog. Phys.} {\bf 66}, 1547 (2003).
  


\bibitem{fauque}
B. Fauqu\'e,  \textit{et~al.}, {\it Phys. Rev. Lett.} {\bf 96}, 197001 (2006).

\bibitem{mook} 
H.A. Mook, \textit{et~al.}, {\it Phys. Rev. B} {\bf 78}, 020506(R) (2008).

\bibitem{li}
Y. Li \textit{et~al.}, {\it Nature} {\bf 455}, 372 (2008).

\bibitem{varmaQCP} 
 M.S. Gr\o nsleth, \textit{et~al.},
{\it Phys. Rev. B} {\bf 79}, 094506 (2009).

\bibitem{varma} 
C.M. Varma, {\it Phys. Rev. B} {\bf 73}, 155113 (2006).

\bibitem{weber} 
 C. Weber \textit{et~al.},
{\it Phys. Rev. Lett.} {\bf 102},  017005 (2009).


\bibitem{yamada} 
K. Yamada \textit{et~al.},
{\it Phys. Rev. B}, {\bf 57}, 6165 (1998).


\bibitem{fujita} M. Fujita \textit{et~al.},
{\it Phys. Rev. B}, {\bf 70}, 104517 (2004).

\bibitem{tranquada-95}
J.M. Tranquada \textit{et~al.},
{\it Nature} {\bf 375}, 561-563 (1995).

\bibitem{tranquada}
 J.M. Tranquada, 
  in {\it Treatise of High Temperature Superconductivity} by J. Robert Schrieffer (2005); arXiv:cond-mat/0512115.

\bibitem{hinkov} 
V. Hinkov \textit{et~al.},
 {\it Science} {\bf 319}, 597 (2008).
 
\bibitem{stripes} 
S.A. Kivelson \textit{et~al.},
{\it Rev. Mod. Phys.} {\bf 75}, 1201 (2003).

\bibitem{julien}  M.-H. Julien,  
 {\it  Physica B}  {\bf 329-333}, 693 (2003).

\bibitem{mitrovic} V.F. Mitrovic \textit{et~al.}, {\it Phys. Rev. B}, {\bf 78}, 014504 (2008).

\bibitem{chang} J. Chang \textit{et~al.}, {\it Phys. Rev. B} {\bf 78}, 104525 (2008).


\bibitem{growth} I.  Tanaka, K. Yamane, and H.  Kojima,  {\it J. Crystal Growth} {\bf 96}, 711 (1989).



\bibitem{oda}
 M. Oda, N.  Momono, and M. Ido, {\it J. of Phys. and Chem. of Sol.} {\bf 65}, 1381 (2004).
 

\bibitem{ido}
M. Ido, N. Momono, and M.  Oda, 
 {\it J. Low. Temp. Phys.} {\bf 117}, 329 (1999).

\bibitem{ong} 
Y. Wang \textit{et~al.},
 {\it Phys. Rev. B} {\bf 64}, 224519 (2001).

  \bibitem{lipscombe} 
O.J. Lipscombe \textit{et~al.},
 {\it Phys. Rev. Lett.} {\bf 102},  167002 (2009).

\bibitem{hirata} 
H. Hiraka \textit{et~al.},
{\it J. Phys. Soc. Jpn}, {\bf 70}, 853–858 (2001).

 \bibitem{gabay} 
 P. Chudzinski, M. Gabay, and T. Giamarchi, {\it Phys. Rev. B} {\bf 78}, 075124 (2008).
 
 \end{thebibliography}
\end{document}